\documentclass{llncs}
\usepackage{amsmath}
\usepackage{graphicx}
\usepackage{subfigure}
\usepackage{ifthen}
\usepackage{enumitem}
\setlist{nolistsep} % compact lists

\def\MST{\mbox{MST}}% \MST(G) is weight of min spanning tree
\def\OPT{\mbox{OPT}}% \OPT(G) is weight of min TSP tour
\def\Left{\mbox{left}}% \Left(v) is left endpoint of interval I_v
\def\In#1{\mbox{in$(#1)$}} % \In{e} is total charge into e
\def\Out#1{\mbox{out$(#1)$}} % \Out{e} is total charge out of e
\def\Net#1{\mbox{net$(#1)$}} % \Net{e} is net charge at e 
\def\xeP{x_{(e,P)}} % charge from e to P

\ifthenelse{\equal{\detokenize{main}}{\jobname}}
{\message{
*** Run pdflatex on 'short.tex' or 'long.tex', not 'main.tex' ***
}\stop}{}

\newboolean{long}
\ifthenelse{\equal{\detokenize{gd12long}}{\jobname}}
{\setboolean{long}{true}}% want long paper
{\setboolean{long}{false}}% want short paper

\title{Light Spanner and Monotone Tree}
\author{Hao-Hsiang Hung}
\institute{Dept.\ of Math \&\ CS, Emory University,
 \textsl{\{hhung2\}@mathcs.emory.edu}}%

\begin{document}
\maketitle

%% Show page numbers in long version
\ifthenelse{\boolean{long}}{
\pagestyle{plain}\thispagestyle{plain}
}{}

\begin{abstract}
In approximation algorithm design, light spanners has applications
in graph-metric problems such as metric TSP (the traveling salesman problem)~\cite{Grigni:2000:ATG:646253.686316} and others~\cite{Berger05approximationschemes}.
We have developed an efficient algorithm in~\cite{pathwidth12} for light spanners in bounded 
pathwidth graphs, based on an intermediate data structure called \emph{monotone tree}.
In this paper, we extended the results to include bounded catwidth graphs.
%Given an edge-weighted graph $G$ and $\epsilon>0$, a
%$(1+\epsilon)$-spanner is a spanning subgraph $G'$ whose shortest path
%distances approximate those of $G$ within a factor of $1+\epsilon$.  
%For $G$ from certain graph families (at least bounded genus and apex
%graphs), we know that \emph{light} spanners exist.  That is, we can
%compute a $(1+\epsilon)$-spanner $G'$ with total edge weight at most a
%constant times the weight of a minimum spanning tree.  This constant
%may depend on $\epsilon$ and the graph family, but not on the
%particular graph $G$ nor on the edge weighting.  The existence of
%light spanners is essential in the design of approximation schemes for
%the metric TSP (the traveling salesman problem) and similar
%graph-metric problems.

%In this paper we make some progress towards the conjecture that light
%spanners exist for every minor-closed graph family: we show that light
%spanners exist for graphs with bounded pathwidth, and they are computed
%by a greedy algorithm. We do this via the intermediate construction of
%light \emph{monotone} spanning trees in such graphs.
%Our results can be easily extended to some graph families with bounded
%treewidth assumption.
\end{abstract}

\section{Introduction}

\emph{Light Spanners}.
Suppose $G$ is a connected undirected graph where each edge $e$ has
length (or weight) $w(e)\geq 0$.  Let $d_G (u, v)$ denote the length
of the shortest path between vertices $u$ and~$v$.  Suppose $G'$ is a
spanning subgraph of $G$, where each edge of $G'$ inherits its weight
from $G$; evidently $d_G(u,v) \leq d_{G'}(u,v)$.  Fix $\epsilon > 0$.
If $d_{G'}(u,v) \leq (1+\epsilon)\cdot d_G (u,v)$ (for all $u,v$),
then we say that $G'$ is a \emph{$(1+\epsilon)$-spanner} of $G$.  In
other words, the metric $d_{G'}$ closely approximates the
metric~$d_G$.

Let $w(G')$ denote the total edge weight of $G'$, and let $\MST(G)$
denote the minimum weight of a spanning tree in $G$.  We are
interested in conditions on $G$ that guarantee the existence of a
$(1+\epsilon)$-spanner $G'$ with bounded $w(G')/\MST(G)$.  More
formally, suppose $\mathcal{G}$ is a family of undirected graphs.  We
say \emph{$\mathcal{G}$ has light spanners} if the following
holds: for every $\epsilon>0$ there is a bound $f(\epsilon)$, so that
for any edge-weighted $G$ from $\mathcal{G}$, $G$ has a
$(1+\epsilon)$-spanner $G'$ with $w(G') \leq f(\epsilon)\cdot\MST(G)$.
Less formally, we say such a $G'$ is a \emph{light spanner} for $G$.
Note $f(\epsilon)$ depends on $\epsilon$ and $\mathcal{G}$, but not on
$G$ or~$w$.

We know that if a graph family has unbounded clique minors, then it
does not have light spanners; just consider a clique with uniform edge
weights.  We conjecture the
converse~\cite{Grigni:2000:ATG:646253.686316}:
\begin{conjecture}\label{conj:main}
Any graph family with a forbidden minor has light spanners.
\end{conjecture}
 
%Our pursuit of this conjecture is guided by the Robertson-Seymour
%theory \cite{DBLP:journals/jct/RobertsonS03a}, which characterizes 
%minor-closed graph families using four elements: bounded genus graphs,
%apices, vortices, and repeated clique-sums.  We already know that if
%$G$ has bounded genus or is an apex graph, then it has light
%spanners~\cite
%{Grigni:2000:ATG:646253.686316,Grigni:2002:LSA:545381.545492}.
%In particular vortices are bounded pathwidth subgraphs, stitched inside
%the faces of a bounded genus graph.
%\newline\newline
%\noindent\emph{Motivation}.
%Conjecture~\ref{conj:main} seems like a natural question, and its
%proof would address the ``main difficulty'' discussed in the
%concluding remarks of Demaine \emph{et
%al.}~\cite{Demaine:2007:AAV:1283383.1283413}, in the general context
%of approximation algorithms on weighted graphs.  As a specific
%motivating problem, we review some results on the \emph{metric TSP},
%the Traveling Salesman Problem with triangle inequality.  (For some
%other problems, see
%\cite{Berger05approximationschemes,DBLP:conf/icalp/BergerG07}.)

Our motivation comes from metricial optimization problems such as 
\emph{metric TSP} and any other problems~\cite{Berger05approximationschemes,DBLP:conf/icalp/BergerG07} need light spanner as an algorithmic tool.
We are given an edge-weighted graph $G$, and we seek a cyclic
order of its vertices with minimum total distance as measured by
$d_G$.  Equivalently, we want a minimum weight cyclic tour in $G$
visiting each vertex at least once.  Let $\OPT(G)$ denote the minimum
tour weight; it is well known that $\MST(G) \leq \OPT(G) \leq
2\cdot\MST(G)$.  We seek an approximation scheme: an algorithm which
takes as inputs the weighted graph $G$ and $\epsilon>0$, and which
outputs a tour with weight at most $(1+\epsilon)\cdot\OPT(G)$.

The problem is MAX SNP-hard~\cite{PY-tsp12-93}, so we consider
approximation schemes where the input graph $G$ is restricted to some
graph family $\mathcal{G}$ (e.g., planar graphs).  We would like a
PTAS (an approximation scheme running in time $O(n^{g(\epsilon)})$,
for some function $g$), or better yet an EPTAS (an approximation
scheme running in time $O(g(\epsilon)\cdot n^c)$, where the constant
$c$ is independent of $\epsilon$).

Suppose $\mathcal{G}$ is a graph family, and that for any $G \in
\mathcal{G}$ we can compute a $(1+\epsilon)$-spanner $G'$ with $w(G')
\leq f(\epsilon)\cdot \MST(G)$.  Then we may attempt to design a PTAS
(or an EPTAS) for the metric TSP on $\mathcal{G}$, as follows:
\begin{enumerate}

\item On input $G$ and $\epsilon$, first compute $G'$, a
  $(1+\epsilon/2)$-spanner of $G$, with weight at most
  $f(\epsilon/2)\cdot \MST(G)$.

\item Choose $\delta = (\epsilon/2)/f(\epsilon/2)$.
Apply some algorithm finding a tour in $G'$ with cost at most
$\OPT(G') + \delta \cdot w(G')$.

\item Return the tour, with cost at most $(1 + \epsilon/2)\cdot\OPT(G)
+ \delta \cdot (f(\epsilon/2)\cdot\MST(G)) \leq
(1+\epsilon)\cdot\OPT(G)$.  (For other metric optimization problems,
it may be less trivial to lift a solution from $G'$ back to $G$.)

\end{enumerate}
Step 2 looks like the original problem, except now we allow an error term
proportional to $w(G')$ instead of $\OPT(G')$.   This
approach has already succeeded for planar graphs~\cite
{DBLP:conf/soda/AroraGKKW98,KleinTSP2005} and bounded genus
graphs~\cite
{Demaine:2007:AAV:1283383.1283413,Grigni:2000:ATG:646253.686316}.

A recent result of Demaine \textit{et
al.}~\cite[Thm.~2]{ContractionMinorFree_STOC2011} implies a PTAS for
metric TSP when $\mathcal{G}$ is \emph{any} graph class with a fixed
forbidden minor.  Since we do not know that $\mathcal{G}$ has light
spanners, for step~1 they substitute a looser result~\cite
{Grigni:2002:LSA:545381.545492}, finding a $(1+\epsilon)$-spanner $G'$
with weight $O((\log n )/\epsilon)\cdot\MST(G)$ (the hidden constant
depending on $\mathcal{G}$).  In step~2 their algorithm runs in time
$2^{O(1/\delta + \log n)}$. Their $1/\delta$ is
$O(w(G')/(\MST(G)\cdot\epsilon))=O((\log n)/\epsilon^2)$, so their
running time is $n^{O(1/\epsilon^2)}$.  If we could compute light
spanners for $\mathcal{G}$, then $\delta$ would improve to something
independent of $n$, and this would yield an EPTAS for metric TSP on
$\mathcal{G}$. (Or alternatively, it would yield an approximation
scheme allowing $\epsilon$ to slowly approach zero, as long as
$1/\delta$ stays $O(\log n)$.)
\newline\newline
\noindent\emph{Our Work}.
A series of previous work of Grigni \emph{et al.} has discovered light spanners in bounded genus graphs~\cite{Grigni:2000:ATG:646253.686316}, apex graphs~\cite{Grigni:2002:LSA:545381.545492},
 and bounded pathwidth graphs~\cite{pathwidth12}.
Obviously, we know a list of graph families with bounded treewidth constraints can also been handled: see \cite[Ch.~7]{DBLP:books/sp/Kloks94} for reference.

We know a intermediate data structure called \emph{monotone tree}
could handle bounded pathwidth graphs well; however, there are simple 
bounded treewidth graphs which do not have any monotone tree~\cite{pathwidth12} 
but has simple charging schemes.
What we are interested in is the strength of the monotone tree.

In this paper, we study the monotone tree in bounded catwidth graphs.
Given a graph $G$ with pathwidth $pw(G)$, treewidth $tw(G)$, and catwidth $catw(G)$,
we know that $tw(G)\leq catw(G)\leq pw(G)$~\cite{smalltree05}.
We handle bounded catwidth graphs based on the fact that for any graph 
$H$ of catwidth $k$, there is a graph of pathwidth $k$ with $(p,q)$-flaps 
(to be defined later) stitched and it contains $H$ as a subgraph.

We prove the following theorem in Section~\ref{sec:mainarg}.
\begin{theorem}\label{thm:main}
Bounded catwidth graphs have light spanners, computable by a greedy
algorithm.
\end{theorem}

The study of catwidth of graphs origins from the memory allocation problems in dynamic programming, simply just because the size of the table required does not grow as fast as bounded treewidth graphs.
Habib et al.~\cite{HPT03} shows that there is a linear time algorithm for recognizing the catwidth of the graph.
Additionally, we know metric TSP in bounded catwidth graphs could be solved in polynomial time~\cite{Cygan11}, based on the fact that bounded catwidth graphs are subclasses of bounded treewidth graphs, and metric TSP is fixed-parameter-tractable in bounded treewidth graphs.
See Section~\ref{sec:further} for some further remarks.

\section{Preliminaries}

\subsection{Charging Schemes}
\label{sec:scheme}

In order to exhibit light spanners in a weight-independent way, we use
\emph{charging schemes}~\cite{Grigni:2002:LSA:545381.545492}.  (We use
the notion called ``0-schemes'' in
\cite{Grigni:2002:LSA:545381.545492}, not the more general
``$\epsilon$-schemes'' required for apex graphs.)  Suppose each edge
of graph $G$ can hold some quantity of \emph{charge}, initially zero.
A \emph{detour} is an edge $e\in E$ and a path $P$ such that $e+P$ is
a simple cycle in $G$.  For each detour $(e,P)$ we introduce a
variable $\xeP\geq0$.  Each $\xeP$ describes a
\emph{charging move}: it subtracts $\xeP$ units of charge from
edge $e$, and adds $\xeP$ units of charge to each edge of $P$.
When $\xeP>0$, we say ``$e$ charges $P$''.
%% we say ``$e$ sends $\xeP$ units of charge to $P$''.

Given graph $G$, a spanning tree $T$, and a number $v$, a
\emph{charging scheme from $G$ to $T$ of value $v$} is an assignment
of nonnegative values to the $\xeP$ variables (i.e., a fractional
sum of detours) meeting the three conditions listed below.  Here 
$\Out{e}$ denotes the total charge subtracted from edge $e$,
$\In{e}$ denotes the total charge added to $e$ (as part of various detour
paths),
and $\Net{e} = \In{e} - \Out{e}$ is the total charge on $e$ after all
the moves are done:
\[
\begin{array}[t]{l@{~~~}rcl@{~~}l}
(1) & \Out{e} & \geq & 1
& \mbox{for all } e \in G-T, \\
(2) &     \Net{e} & \leq & 0 
%\epsilon\cdot(\Out{e}-1),
& \mbox{for all } e \in G-T, \\
(3) &     \Net{e} & \leq & v
% v + \epsilon\cdot \Out{e},
& \mbox{for all } e \in T.
\end{array}
\]
Note ``$e \in G-T$'' means $e$ is an edge of $G$ but not $T$.
As we'll see in Theorem~\ref{thm:greedy}, charging schemes imply light
spanners.
\begin{definition} 
An \emph{acyclic scheme} is a charging scheme with two additional
properties:
\begin{description}
\item[\textnormal{(4)}] If edge $e$ charges some path, then $e\in
G-T$.
\item[\textnormal{(5)}] There is an ordering of the edges such that whenever
edge $e_1$ charges a path containing edge~$e_2$, $e_1$ precedes~$e_2$.
\end{description}
\end{definition}
For example, planar graphs have integral acyclic
schemes of value $v=2$~\cite{Althofer:1993:SSW:156252.156258}.
%% FALSE, these schemes are *sums* of O(g) acylic schemes:
%%   and genus $g$ graphs have acyclic charging schemes of value
%%   $O(g)$~\cite{Grigni:2000:ATG:646253.686316}.
\begin{definition}
Suppose we have detours $(e_1,P_1)$ and $(e_2,P_2)$, with $e_2\in P_1$
and $e_1\not\in P_2$.  Their \emph{shortcut} is the detour $(e_1,P')$,
where $P'$ is the path derived from $P_1$ by replacing $e_2$ with
$P_2$, and then reducing that walk to a simple path.
\end{definition}

\begin{lemma}\label{lem:shortcut} 
Suppose we have an acyclic scheme of value $v$ from $G$ to $T$, and
an edge $e$ in $G-T$.  Then there is an acyclic scheme of value $v$
from $G-e$ to $T$.
\end{lemma}

\begin{proof}
Let $e_2=e$.  While $\In{e_2}$ is positive, we find some $e_1$ charging
a path $P_1$ containing $e_2$.  Since $\Net{e_2}
\leq 0$, $e_2$ also charges some path $P_2$.  $P_2$
cannot contain $e_1$, since the scheme is acyclic.  Let
$\alpha=\min(x_{(e_1,P_1)}, x_{(e_2,P_2)})$.  Now reduce both
$x_{(e_1,P_1)}$ and $x_{(e_2,P_2)}$ by $\alpha$, and increase
$x_{(e_1,P')}$ (their shortcut) by $\alpha$.  After this change all the
conditions are still satisfied, except possibly for condition (1) at $e_2$.
Repeat until $\In{e_2}$ reaches zero.  Finally remove $e_2$ and any
remaining charges out of~$e_2$.  \qed
\end{proof}

\begin{theorem}
\label{thm:greedy}
%% Recently: removed G^* / G distinction, no need.
Suppose $G$ is a graph with spanning tree $T$, and we have an acyclic
scheme from $G$ to $T$ of value $v$.  Then for any $\epsilon > 0$, and
for any non-negative edge-weighting $w$ on $G$, a simple greedy
algorithm finds a $(1+\epsilon)$-spanner $G'$ in $G$ containing $T$,
with total weight $w(G')\leq(1+v/\epsilon)\cdot w(T)$.
\end{theorem}
We use the following greedy algorithm of Alth\"{o}fer 
\emph{et al.}~\cite{Althofer:1993:SSW:156252.156258},
modified to force the edges of $T$ into~$G'$:
\begin{quote}
\begin{tabbing}
xxx \= xxx \= xxx \= \kill
Spanner($G$, $T$, $1+\epsilon$): \\
\> $G' = T$ \\
\> for each edge $e\in G-T$, in non-decreasing $w(e)$ order \\
\>\> if $(1+\epsilon)\cdot w(e) < d_{G'}(e)$ then \\
\>\>\> add edge $e$ to $G'$\\
\>return $G'$
\end{tabbing}
\end{quote}
The proof of Theorem~\ref{thm:greedy} is a variant of previous 
arguments by LP duality~\cite
{Grigni:2000:ATG:646253.686316,Grigni:2002:LSA:545381.545492},
\ifthenelse{\boolean{long}}{
for completeness we sketch it here.
\begin{proof}
Since $G'$ is computed by the greedy algorithm, it is clearly a
$(1+\epsilon)$-spanner of $G$ containing $T$; the issue is to bound
its weight $w(G')$.  
By Lemma~\ref{lem:shortcut} we have an acyclic scheme from $G'$ to
$T$ of value~$v$.  

Consider a detour $(e,P)$ in $G'$ with $e\not\in
T$. We claim $(1+\epsilon)\cdot w(e) < w(P)$ (to see this, compare $e$
with the last edge inserted by the algorithm on the cycle $e+P$). 
Multiply through by $\xeP$ and we have this:
\begin{eqnarray*}
\xeP \cdot \epsilon \cdot w(e) & \leq & \xeP \cdot (w(P)-w(e))
\end{eqnarray*}
When $e\in T$ this is still valid, since $\xeP=0$.  Now sum over all
detours $(e,P)$:
\begin{eqnarray*}
\sum_{(e,P)} \xeP \cdot \epsilon \cdot w(e)
& \leq &
\sum_{(e,P)} \xeP \cdot (w(P)-w(e)) \\
\epsilon \cdot \sum_{e\in G'} w(e)\cdot \Out{e}
& \leq &
\sum_{e\in G'} w(e) \cdot \Net{e} \\
\epsilon \cdot w(G'-T) & \leq & v \cdot w(T)
\end{eqnarray*}
So $w(G') = w(T)+ w(G'-T) \leq (1+v/\epsilon)\cdot w(T)$.
\qed
\end{proof}
}{
%% Short version:
based on Lemma~\ref{lem:shortcut} that we have an acyclic scheme from $G'$ to
$T$ of value~$v$. We omit the detail here. 
}

\subsection{Treewidth, Pathwidth, Catwidth and Monotone Trees}
\label{def:bpmt}
Suppose $G=(V,E)$ is a graph, $T$ is a tree, and
$\mathcal{B}={(B_i)}_{i\in T}$ is a collection of subsets of $V$ (bags) indexed
by vertices $i$ (identical copies) in $T$.  We call the pair $(T, \mathcal{B})$ a
\emph{tree decomposition} of $G$ if the following conditions hold: (1)
${\bigcup}_{i\in T} B_i = V$; (2) for every $e=\{u, v\}\in E$, there
is at least one bag $B_i$ where $u\in B_i$ and $v\in B_i$; (3) for
every $v\in V$, the collection of $B_i$ containing $v$ is connected
(an interval) in $T$.  The \emph{treewidth} of the decomposition is
the maximum bag size minus one, and the treewidth of $G$ is the
minimum treewidth of any tree decomposition of $G$.

If $T$ is a path (that is, $(P,\mathcal{B})$ instead of $(T, \mathcal{B})$ where 
$P$ stands for a path), then we call it a path decomposition of $G$ (and of course,
the \emph{pathwidth} of $G$ is the minimum width of any path decomposition of $G$).
Additionally, if $T$ is a caterpillar (a path with added leaves, $(C,\mathcal{B})$ where 
$C$ stands for a caterpillar) then we call it a caterpillar decomposition of $G$, with 
`\emph{catwidth}' defined in a similar way.

How to represent a bounded pathwidth graph?
Given $(P, \mathcal{B})$, we may lay out $P$ on the line, and regard
$G$ as a subgraph of an interval graph.  That is, for each vertex $v$
we have a line interval $I_v$ (corresponding to an interval in $P$),
and we have $I_u \bigcap I_v\neq\emptyset$ whenever $\{u, v\}\in E$,
and at most $k+1$ intervals overlap at any point of the line.  For
convenience we may eliminate ties, so that all the interval endpoints
are distinct.  In particular, let $\Left(v)$ denote the leftmost point
of $I_v$.   

%If $P$ is not only a path but also a tree, then we call $(P, \mathcal{B})$ a
%\emph{tree decomposition} of $G$ if the above three conditions hold
%(the only difference of the definition is the topology of $P$).
%We call a tree decomposition \emph{nice}\cite{DBLP:books/sp/Kloks94}
% if it fulfills the following additional conditions: (1) $P$ is a binary rooted 
%tree; (2) if $i\in P$ has two children $i_1$ and $i_2$, then 
%$B_i = B_{i_1} = B_{i_2}$; (3) if $i\in P$ has only one child $i_1$, then 
%either $B_i\subset B_{i_1}$ and $|B_{i_1} - B_i| = 1$, or 
%$B_{i_1}\subset B_i$ and $|B_i - B_{i_1}| = 1$.

Next, we define the immediate data structure for helping the analysis later.
Suppose $T$ is a rooted tree in $G$ (here we abuse the notation of $T$ 
a little bit without confusion).  We say $T$ is \emph{monotone} if
for every vertex $v$ in $T$ with parent $p$, we have $\Left(p) <
\Left(v)$.  When $T$ is a path rooted at an endpoint, we say it is a
\emph{monotone path}.  In particular if $T$ is a monotone spanning
tree in $G$, then from any vertex $v$, we can find a monotone path in
$T$ from $v$ to the root of $T$ (the vertex with the leftmost
interval).  For this process, it is convenient to imagine that edges
connect intervals at their leftmost intersection point.

\subsection{$(p,q)$-flap}

We first introduce parameters to the graph structure as follows.
A $k$-clique is a clique of size $k$.
A $k$-leaf is the union of one external vertex $v$ attached to a $k$-clique,
with additional edges connecting $v$ to all the vertices of this clique.
A $k$-tree is a graph which can be constructed as follows: starting from a 
$k$-clique, repeatedly adding $k$-leaves to the existent graph.
A $k$-path is a graph which can be constructed by repeatedly adding
$k$-leaves to a proper $k$-path such that the neighbor of each $k$-leaf is 
a separator of the original proper $k$-path.

A $(p,q)$-flap is a partition $(P,Q)$ of vertices from a clique (in a graph $G$) of size $p+q$
such that (1) $|P|=p$; (2) $|Q|=q$; and (3) $P$ separates $Q$ from the rest of $G$.
We also define the attachment of a $(p,q)$-flap to a graph $G$ as follows: first identify a clique $P$ of size 
$p$ in $G$, then add an external clique $Q$ of size $q$ with additional edges $(u,v)$ between all $u\in P$ 
and all $v\in Q$.
Note that a $k$-leaf is actually a $(k,1)$-flap.

A $k$-caterpillar is a $k$-path with attached $(p,q)$-flaps such that $p+q=k+1$, in fact, it is also a $k$-tree.
It is known that a graph $G$ has catwidth at most $k$ if and only if $G$ is a partial $k$-caterpillar~\cite{smalltree05}.

\section{Surgery for $k$-caterpillars}
\label{sec:reduction}

We are given $\epsilon>0$, a connected edge-weighted graph $G$ with
$n$ vertices, and an interval representation $\{I_v\}$ of $G$ with
pathwidth $k$, with $(p,q)$-flaps attached (that is, a $k$-caterpillar).  We want to find a $(1+\epsilon)$-spanner $G'$ in $G$
of low weight.  First we apply some reductions to simplify $G$ in the following subsections.
In each of the following subsections, the first paragraph discusses the $k$-path (of bounded pathwidth) alone,
and the second paragraph discusses the attachment of the flaps.

\subsection{Nice Path Decomposition}.
\label{sec:nice}

We first assume that each pair of consecutive bags (as vertex
sets) differ by only one vertex.  This can be enforced by an argument
similar to the construction of \emph{nice} tree-decompositions
\cite{DBLP:books/sp/Kloks94}: if two consecutive bags differ on $m\geq
2$ vertices, we introduce $m-1$ intermediate bags, in such a way that
each pair differs on only one vertex, and we do not increase the
maximum bag size. This does not modify $G$ at all.

Because a $(p,q)$-flap is attached to a $q$-clique of the $k$-path, and by nice path decomposition
there should be one bag containing all vertices of a $q$-clique (if the last vertex added in one bag but
some vertices of the clique are removed then it is impossible to form a $q$-clique), therefore we simply
attach each one of them to a corresponding bag (one bag may have many flaps attached).

\subsection{Bounded Degree Assumption}.
\label{sec:degree}

We may assume each vertex appears in $O(k)$ bags, and so the
maximum degree of $G$ is $O(k)$.  To enforce this, we copy the bags of $G$
from left to right.  After each group of $k$ original bags, ending
with a bag $B$, we insert $|B|$ ``replacer'' bags, each of which
replaces one vertex $v\in B$ with a copy $v'$, connected to $v$
by an edge of length zero. This ends with a bag $B'$, where every
vertex $v \in B$ has been replaced by a copy $v'\in B'$.
\ifthenelse{\boolean{long}}{See Figure~\ref{fig:nicepath}.}{}
We continue in this way (using the copies in place of the originals)
across the entire path decomposition.  If we aren't careful the
pathwidth may increase by one, but this does not matter for our
asymptotic results.  The original graph is obtained by contracting a
set $S$ of weight-zero edges in the modified graph.  So given a
spanner $G'$ in this modified graph, we may contract $S$ in $G'\cup S$
to recover a spanner (of no greater weight) in the original.

The $O(k)$ maximum degree assumption does not introduce any difficulty to the 
representation of the flaps at all because
the degree of each attached vertex is also bounded by $k$.
Therefore even if a vertex appears in multiple $q$-cliques and has to be separated into different
identical copies by this assumption, we can attach these $(p,q)$-leaves individually according to which $q$-clique they attach to.

\subsection{Completion Assumption}
\label{sec:complete}

We may assume that $G$ is \emph{completed}; that is, it contains
all edges allowed by its overlapping intervals (in other words: we
have a clique in each bag, $G$ is an interval graph).  For each absent
edge $e=\{u,v\}$, we simply add it with weight $w(e)$ equal to the
shortest path length $d_G(u,v)$.  This does not change $d_G$ at all.
Given a spanner $G'$ in the completed graph, we recover a spanner in
the original graph by replacing each completion edge by the
corresponding shortest path, and apparently this assumption provides
an upper bound of the charges towards the edges in $T$.

\ifthenelse{\boolean{long}}{
\begin{figure}[ht]
\begin{center}
{\includegraphics[scale=0.1]{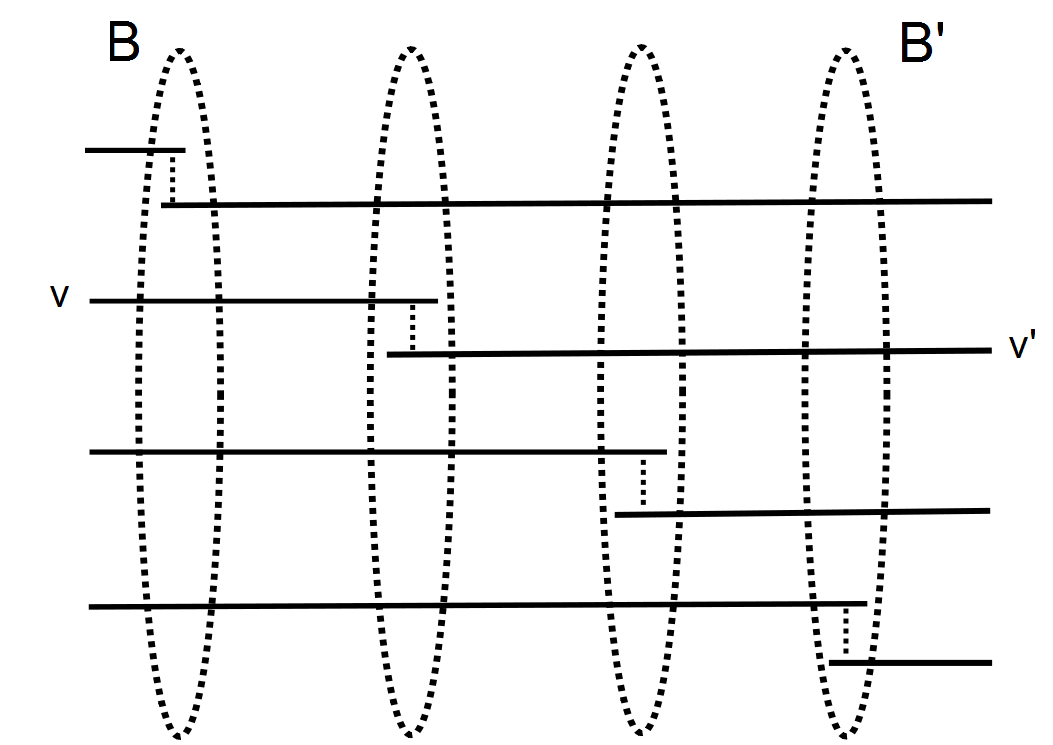}}
\caption{Each vertex $v$ in bag $B$ is replaced by $v'$ in bag $B'$.}
\label{fig:nicepath}
\end{center}
\end{figure}
}{}

Since we attach each flap to a really `complete' subgraph of the original
graph, we can limit the charging moves introduced by the new additional $p\cdot q$ edges 
within the subgraph induced by these $k+1$ vertices of the clique.

\section{Our Approach}
\label{sec:mainarg}

We start by proving the main theorem as follows (assume all the reductions
in Section~\ref{sec:reduction} are applied).

\begin{proof}[of Theorem~\ref{thm:main}]
We assume all the above reductions have been applied: the input graph
$G$ is completed (an interval graph), each bag introduces at most one
vertex, and each vertex has degree $O(k)$.

By Lemma~\ref{lem:monoT} (below), we compute a monotone spanning tree
$T$ with $w(T)=O(k^2)\cdot \MST(G)$.
We handle the charges from the $k$-paths and the flaps individually.
By Lemma~\ref{lem:T2scheme} (below), we exhibit an acyclic charging
scheme from $G$ to $T$ of value $v=O(k)$.
By Lemma~\ref{lem:flapcharge} (below), we deal with charges from the flaps 
to $T$ of value $v=O(k)$.
Finally we apply the greedy algorithm, which computes
a ($1+\epsilon$)-spanner $G'$.
By Theorem~\ref{thm:greedy}, $w(G') \leq (1+v/\epsilon)\cdot w(T) =
O(k^3/\epsilon)\cdot \MST(G)$.
\qed
\end{proof}

\begin{lemma}\label{lem:monoT}Given a weighted, bounded catwidth
graph $G$ is completed under shortest path metric. There is a monotone
 spanning tree $T$ inside of the $k$-path part with $w(T)\leq O(k^2)\cdot \MST(G)$. 
\end{lemma}

\begin{proof}
Choose a minimum spanning tree $T^*$, so $w(T^*)=\MST(G)$.
Let $I_l$ and $I_r$ be the leftmost and rightmost intervals.
Let $P_1$ be a shortest path from $I_l$ to $I_r$; since $G$ is completed,
we may assume $P_1$ is monotone, as in Figure~\ref{fig:tree1}.
Note $w(P_1)\leq w(T^*)$.

Consider the components ${T_1}^*, {T_2}^*, ..., {T_m}^*$ of $T^* - V(P_1)$.
Let $e_i$ be an edge connecting the leftmost point of $T_i^\ast$ to a
vertex of $P_1$ (it exists by completion).
For each ${T_i}^*$, we recursively compute a monotone spanning tree $T_i$
of $G[V({T_i}^*)]$. Finally, $T = P_1 \cup \bigcup_i (T_i\cup e_i)$.

We also have to consider the vertices from the $(p,q)$-flaps.
Given a $(p,q)$-flap, we consider the edges of MST of $G$ in the flap.
The edges might be divided into $j$ components ($1\leq j\leq n$, and the connection paths are at $G-P-Q$),
and we denote them as $c_1,\cdots,c_j$.
Within each component if the removal of its $Q$ side will partition it into sub-components of MST then
we connect ends of the detour paths of the MST induced in its $P$ side, with the weights of these completion edges setup as the weights of the shortest path distances inside of the MST.

It is clear that $T$ is monotone (although we allow some twist paths inside of the same bag by flaps), but we must account for the total
weight of $w(T)$.  For each component ${T_i}^*$, let $f_i$ be an edge
of $T^*$ connecting ${T_i}^*$ to $P_1$ (there must be at least one).
By triangle inequality, we see $w(e_i)$ is at most $w({T_i}^*) +
w(f_i) + w(P_{1,i})$, where $P_{1,i}$ is a sub-path of $P_1$ from
the endpoint of $e_i$ to the endpoint of $f_i$.  Note the $f_i$'s and
$T^*_i$'s are disjoint parts of $T^*$, but the sub-paths may overlap inside
$P_1$.

\begin{figure}[ht]
\begin{center}
%\vspace*{-1cm} % cannot! tree1.png has an opaque white background? checked, opacity = 100%
{\includegraphics[width=5cm]{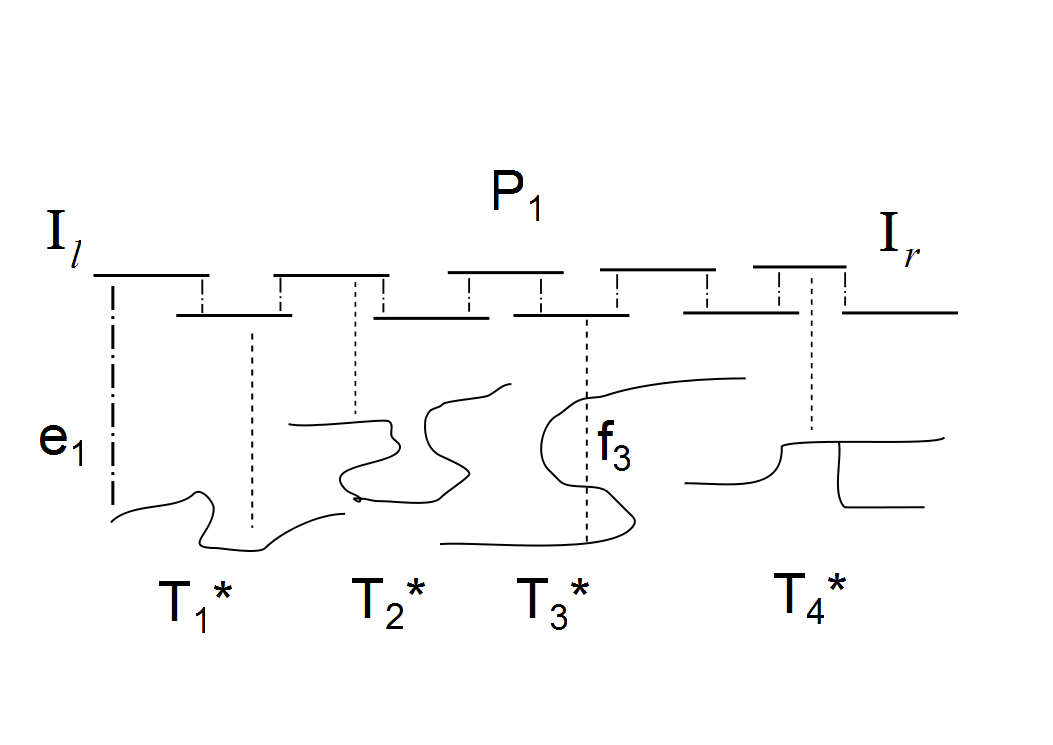}}
\caption{$P_1$ and the ${T_i}^*$ subtrees. Each 
$f_i$ in $T^*$ is replaced by an $e_i$ in $T$.}
\label{fig:tree1}
\end{center}
\end{figure}

%Note for each branch of the MST stretched into the a flap ${T_i}^*$, its $e_i=f_i$ so we do not have to consider the detour path.
%We analyze $w(T)$ as follows.

An edge $e\in P_1$ appears in at most $k-1$ of the
$P_{1,i}$ sub-paths, since each sub-path witnesses another vertex (from
$T^*_i$) that must appear in the bag with~$e$.  So $\sum_i
w(e_i)\leq\sum_i [w(f_i)+w({T_i}^*)+w(P_{1,i})]\leq
w(T^*)+(k-1)w(P_1)\leq k\cdot w(T^*)$.  Since $w(T) \leq O(k\cdot
w(T^*)) + \sum_i w(T_i)$ and $\sum_i w({T_i}^*) \leq w(T^*)$, a simple
depth-$k$ recursion finishes our bound.

\qed
\end{proof}

\noindent\emph{Remark:} we do not have to compute $T$ as in
Lemma~\ref{lem:monoT}; it suffices to use any light enough monotone
spanning tree.  A natural choice is to let $T$ be the \emph{lightest}
monotone spanning tree, which we compute as follows.  Start with just
the root (in the leftmost bag), and grow the tree in a left-to-right
scan of the bags: each time a bag $B$ introduces a new vertex $v$, add
an edge connecting $v$ to its nearest neighbor in $B$ (which is
already in~$T$).

In the completed $G$ (please refer to~\ref{sec:reduction}), a 
\emph{triangle move} is a charging move where a non-tree edge $e$ charges a path $P$ of length two, where at most
one edge of $P$ is not in $T$.  We now define $T^{(2)}$, a graph whose
edges represent triangle moves.  Each vertex $jk$ of $T^{(2)}$
corresponds to an edge $\{j,k\}$ in $G$.  We also represent the vertex
$jk$ by the interval $I_{jk} = I_j \cap I_k$.  To define the edges of
$T^{(2)}$, we first define a \emph{parent} for each vertex $jk$.  If
$\{j,k\}$ is an edge of $T$, then $jk$ has no parent.  Otherwise,
suppose $\Left(j)<\Left(k)$ (else swap them), and let $i$ be the
parent of $k$ in $T$; $i$ must exist since $k$ is not the root.  Note
$\{i,j,k\}$ is a triangle in $G$.  Now we say the parent of $jk$ is
$ij$, and we add the edge $\{ij,jk\}$ in $T^{(2)}$.  Note
$\Left(ij)<\Left(jk)$, so these parent links are acyclic. Thus
$T^{(2)}$ is a forest, with each component rooted at a vertex
corresponding to an edge of $T$.  
%Figure~\ref{fig:pages} illustrates a simple monotone tree $T$ and its forest $T^{(2)}$.

%\begin{figure}[ht]
%\begin{center}
%\mbox{
%\subfigure[a monotone tree $T$]{\includegraphics[width=5cm]{tree2.png}}
%\subfigure[$T^{(2)}$ produced from $T$]{\includegraphics[width=5cm]{tree4.png}}
%}
%% Eliminated (c),(d), since our T^{(2)} construction
%% is not well defined when T is not monotone, e.g. the set {i,j,k}
%% might not be a triangle.
%Figures~\ref{fig:pages}(c) and \ref{fig:pages}(d) show a
%non-monotone $T$ and its $T^{(2)}$, which is ``broken'' since the
%non-tree edge $ae$ is not connected to a tree edge.
%\mbox{
%\subfigure[non-monotone tree $T$]{\includegraphics[width=5cm]{sshape1.png}}
%\subfigure[$T^{(2)}$ produced from $T$ in (c), note disconnected $ae$]{\includegraphics[width=5cm]{sshape3.png}}
%}
%\mbox{
%\subfigure[The interval representation of a graph $G$ and the charging between edges.]{\includegraphics[width=5cm]{tri1.png}}
%\subfigure[A drawing of $G$ with triangulation.]{\includegraphics[width=5cm]{tri2.png}}
%}
%\caption{Horizontal lines are intervals, dashed verticals are edges of $T$.}
%\label{fig:pages}
%\end{center}
%\end{figure}

\begin{lemma}\label{lem:T2scheme} Suppose a weighted, bounded catwidth
graph $G$ is completed under shortest path metric. There is an acyclic 
charging scheme from the $k$-path of $G$ to $T$ of value $O(k)$.
\end{lemma}

%\begin{figure}[ht]
%\begin{center}
%\mbox{%
%\subfigure[interval representation]{\includegraphics[width=6cm]{tri1.png}}
%\subfigure[standard graph drawing]{\includegraphics[width=6cm]{tri2.png}}}
%\caption{A chain of triangle moves (arrows), darker edges are tree edges.}
%\label{fig:trees}
%\end{center}
%\end{figure}

\begin{proof}
Recall $T^{(2)}$ is a forest.  Fix a component $C$ of $T^{(2)}$; it is
a tree, rooted at a vertex $r$ corresponding to an edge of $T$, and
that is the only such vertex in $C$.  Consider a directed Euler tour
of $C$, traversing each edge twice.  Delete each tour edge out of $r$,
so we get a list of directed paths, each of the form
\[
e_1 \rightarrow e_2 \rightarrow \cdots e_m \rightarrow r
\]
where each vertex $e_i$ corresponds to some edge of $G-T$.  Since $C$
is a tree, these paths are vertex disjoint (except at $r$).  However,
a vertex may appear more than once on the same path; call an
appearance $e_i$ a \emph{repeat} if the same vertex appeared earlier
on the path.  Let $\mathcal{P}$ be the collection of all these paths,
from all components of $T^{(2)}$.

We now propose a charging scheme (which fails to be acyclic).  Recall
how we constructed edges in $T^{(2)}$: we connect each vertex $jk$
(corresponding to an edge of $G-T$) to its parent $ij$.  If a path in
$\mathcal{P}$ traverses this edge in the direction $jk \rightarrow
ij$, we add the triangle move where edge $\{j,k\}$ charges one unit
to path $j-i-k$.  If a path traverses this edge in the other
direction $ij \rightarrow jk$ (so $ij$ is not a tree edge), we
add the triangle move where edge $\{i,j\}$ charges one unit to path $i-k-j$.
In either direction, the tree edge $\{i,k\}$ is charged.

We now verify the proposed charging scheme with the properties in Section
~\ref{sec:scheme}.
For an edge $e \in G-T$, the corresponding vertex appears at least
once on a path, and it has at least as many out-edges as in-edges, so
our proposed scheme has properties (1) and (2).  For an edge $e \in
T$, we must bound the number of times it is charged.  Since $G$ has
maximum degree $O(k)$, $e$ appears in $O(k)$ distinct triangles, and it is
charged at most twice per triangle (this includes the charges it
receives in its role as $r$).  So if we choose $v=O(k)$, we have property (3).
Also there are no charges out of tree edges, so we have property (4).

However, this charging scheme does not have property (5); if a vertex
(corresponding to an edge $e \in G-T$) has a repeat appearance on its
path, then there is no consistent way to order the edges.  To fix
this, we eliminate all ``repeat'' appearances using shortcuts.
That is, whenever we have a sequence $e_1 \rightarrow e_2 
\rightarrow e_3$ where $e_2$ is a repeat, we shortcut out $e_2$.
Note that this can be repeated.  For example if we have a sequence
$e_1 \rightarrow e_2 \rightarrow e_3 \rightarrow e_4 \rightarrow e_5$,
corresponding to four triangle moves, it is possible to shortcut out
$e_2,e_3,e_4$ (in any order), and the result is a single charge from
$e_1$ to a path containing $e_5$ (the rest of the charged path is all
tree edges).  After eliminating all repeats by shortcuts, we get the
desired acyclic scheme.
\qed
\end{proof}

\begin{lemma}\label{lem:flapcharge} Suppose a weighted, bounded catwidth
graph $G$ is completed under shortest path metric. There is an acyclic 
charging scheme from $(p,q)$-flaps to $T$ of value $O(k)$.
\end{lemma}

\begin{proof}
Inside of each MST component in a flap (in particular the $Q$ part), the charges 
are localized with value $O(k)$ because the detour paths are in the flap, too. 
In addition, we have to consider influx charges from the $k$-path, and 
by Lemma~\ref{lem:T2scheme} we know it introduced charges of value $O(k)$.

For edges crossing different MST components in a flap, the outflux charges
towards the $k$-path are also of value $O(k)$.
Consider two components $c_i$ and $c_j$ ($i\neq j$, $|c_i|\leq k$, and $|c_j|\leq k$).
For each $u\in c_i$ we group up all $|c_j|$ edges (with one end $u$) with an arbitrary
acyclic charging ordering, and let the last edge charge toward the $k$-path.
Since $|c_i|\leq k$, the value is $O(k)$.  
\end{proof}

\section{Conclusion and Further Work}
\label{sec:further}

In this paper, we extended the technique that we used for finding
light spanners in bounded pathwidth graphs towards bounded catwidth graphs,
with additional charges from `local' structures called $(p,q)$-flaps.
However, a different approach might require in the progress towards Conjecture~\ref{conj:main}.

We have discussed the major difficulty in handling bounded treewidth
graphs in~\cite{pathwidth12}: we have no control over the MST topology
inside of the bounded treewidth graphs, and a trimming might introduce
too heavy overload (relative to the total weight of the MST).

We are investigating known subclasses of bounded treewidth graphs,
and considering introduce some constraints such as bounded degree of the original graphs,
the diameter of the decomposition tree, etc.

\ifthenelse{\boolean{long}}{\newpage}{}

\bibliographystyle{plain}
\bibliography{ref}

\begin{thebibliography}{10}

\bibitem{Althofer:1993:SSW:156252.156258}
Ingo Alth\"{o}fer, Gautam Das, David Dobkin, Deborah Joseph, and Jos\'{e}
  Soares.
\newblock On sparse spanners of weighted graphs.
\newblock {\em Discrete Comput. Geom.}, 9:81--100, January 1993.

\bibitem{DBLP:conf/soda/AroraGKKW98}
Sanjeev Arora, Michelangelo Grigni, David~R. Karger, Philip~N. Klein, and
  Andrzej Woloszyn.
\newblock A polynomial-time approximation scheme for weighted planar graph
  {TSP}.
\newblock In {\em SODA}, pages 33--41, 1998.

\bibitem{Berger05approximationschemes}
Andr{\'e} Berger, Artur Czumaj, Michelangelo Grigni, and Hairong Zhao.
\newblock Approximation schemes for minimum 2-connected spanning subgraphs in
  weighted planar graphs.
\newblock In {\em in Proc. of ESA}, pages 472--483, 2005.

\bibitem{DBLP:conf/icalp/BergerG07}
Andr{\'e} Berger and Michelangelo Grigni.
\newblock Minimum weight 2-edge-connected spanning subgraphs in planar graphs.
\newblock In Lars Arge, Christian Cachin, Tomasz Jurdzinski, and Andrzej
  Tarlecki, editors, {\em ICALP}, volume 4596 of {\em Lecture Notes in Computer
  Science}, pages 90--101. Springer, 2007.

\bibitem{Cygan11}
Marek Cygan, Jesper Nederlof, Marcin Pilipczuk, Michal Pilipczuk, Johan M.~M.
  van Rooij, and Jakub~Onufry Wojtaszczyk.
\newblock Solving connectivity problems parameterized by treewidth in single
  exponential time.
\newblock In Rafail Ostrovsky, editor, {\em FOCS}, pages 150--159. IEEE, 2011.

\bibitem{ContractionMinorFree_STOC2011}
Erik~D. Demaine, MohammadTaghi Hajiaghayi, and Ken ichi Kawarabayashi.
\newblock Contraction decomposition in $h$-minor-free graphs and algorithmic
  applications.
\newblock In {\em Proceedings of the 43rd ACM Symposium on Theory of Computing
  (STOC 2011)}, page to appear, June 6--8 2011.

\bibitem{Demaine:2007:AAV:1283383.1283413}
Erik~D. Demaine, MohammadTaghi Hajiaghayi, and Bojan Mohar.
\newblock Approximation algorithms via contraction decomposition.
\newblock In {\em Proceedings of the eighteenth annual ACM-SIAM symposium on
  Discrete algorithms}, SODA '07, pages 278--287, Philadelphia, PA, USA, 2007.
  Society for Industrial and Applied Mathematics.

\bibitem{Grigni:2000:ATG:646253.686316}
Michelangelo Grigni.
\newblock Approximate {TSP} in graphs with forbidden minors.
\newblock In {\em Proceedings of the 27th International Colloquium on Automata,
  Languages and Programming}, ICALP '00, pages 869--877, London, UK, 2000.
  Springer-Verlag.

\bibitem{pathwidth12}
Michelangelo Grigni and Hao-Hsiang Hung.
\newblock Finding light spanners in bounded pathwidth graphs.
\newblock {\em CoRR}, abs/1104.4669, 2011.

\bibitem{Grigni:2002:LSA:545381.545492}
Michelangelo Grigni and Papa Sissokho.
\newblock Light spanners and approximate {TSP} in weighted graphs with
  forbidden minors.
\newblock In {\em Proceedings of the thirteenth annual ACM-SIAM symposium on
  Discrete algorithms}, SODA '02, pages 852--857, Philadelphia, PA, USA, 2002.
  Society for Industrial and Applied Mathematics.

\bibitem{HPT03}
M.~Habib, C.~Paul, and J.A. Telle.
\newblock A linear-time algorithm for recognition of catval graphs.
\newblock Technical Report RR-LIRMM-03004, LIRMM, Université de Montpellier 2,
  2003.

\bibitem{KleinTSP2005}
Philip~N. Klein.
\newblock A linear-time approximation scheme for tsp for planar weighted
  graphs.
\newblock In {\em Proceedings, 46th IEEE Symposium on Foundations of Computer
  Science}, pages 146--155, 2005.

\bibitem{DBLP:books/sp/Kloks94}
Ton Kloks.
\newblock {\em Treewidth, Computations and Approximations}, volume 842 of {\em
  Lecture Notes in Computer Science}.
\newblock Springer, 1994.

\bibitem{PY-tsp12-93}
Christos~H. Papadimitriou and M.~Yannakakis.
\newblock The {T}raveling {S}alesman {P}roblem with distances one and two.
\newblock {\em Mathematics of Operations Research}, 18:1--11, 1993.

\bibitem{smalltree05}
Jan~Arne Telle.
\newblock Tree-decompositions of small pathwidth.
\newblock {\em Discrete Appl. Math.}, 145(2):210--218, January 2005.

\end{thebibliography}

\end{document}